\documentclass[aps,twocolumn,superscriptaddress,floatfix,prl]{revtex4-2}

\usepackage{lipsum}
\usepackage{graphicx}
\usepackage{amsmath,amssymb}
\usepackage{braket}
\usepackage{mathtools}
\usepackage{hyperref}
\usepackage{multirow}
\usepackage{color}
\usepackage[normalem]{ulem}
\usepackage{amsfonts}
\usepackage{float}
\usepackage{pdfpages} 
\makeatletter
\usepackage{subfigure}
\usepackage{soul, xcolor}

\setstcolor{red}

\def\bra#1{\mathinner{\langle{#1}|}}
\def\ket#1{\mathinner{|{#1}\rangle}}

\renewcommand{\l}{\left(}
\renewcommand{\r}{\right)}
\newcommand{\lb}{\left[}
\newcommand{\rb}{\right]}

\newcommand{\mc}[1]{\mathcal{\mathrel{#1}}}

\def\beq{\begin{equation}}
\def\eeq{\end{equation}}
\def\bea{\begin{eqnarray}}
\def\eea{\end{eqnarray}}

\AtBeginDocument{\let\LS@rot\@undefined}
\makeatother

\begin{document}

\title{Spectral gaps of local quantum channels in the weak-dissipation limit}

\author{J.~Alexander~Jacoby}
\affiliation{Department of Physics, Princeton University, Princeton, New Jersey 08544, USA}

\author{David~A.~Huse}
\affiliation{Department of Physics, Princeton University, Princeton, New Jersey 08544, USA}

\author{Sarang~Gopalakrishnan}
\affiliation{Department of Electrical and Computer Engineering, Princeton University, Princeton NJ 08544, USA}

\date{\today}

\begin{abstract}
    We consider the dynamics of generic chaotic quantum many-body systems with no conservation laws, subject to weak bulk dissipation. It was recently observed [T. Mori, arXiv:2311.10304] that the generator of these dissipative dynamics, a quantum channel $\mathcal{E}$, retains a nonzero gap as the dissipation strength $\gamma \to 0$ if the thermodynamic limit is taken first. We use a hydrodynamic description of operator spreading in the presence of dissipation to estimate the gap of $\mathcal{E}$ as $\gamma \to 0$; to calculate the operator-size distribution of the low-lying eigenmodes of $\mathcal{E}$; and to relate the gap to the long-time decay rates of autocorrelation functions under unitary dynamics. We provide a microscopic derivation of this hydrodynamic perspective for random unitary circuits. We argue that the gap in the $\gamma \to 0$ limit can change nonanalytically  as one tunes the parameters of the unitary dynamics.
\end{abstract}

\maketitle

A chaotic many-body system ``acts as its own bath''~\cite{basko2006metal} in that the dynamics of a subsystem of interest can be described by treating the rest of the system as a bath. Coupling a chaotic system weakly to an external bath does not change the dynamics of conventional local observables. However, it drastically reduces the computational complexity of classical simulations of quantum many-body dynamics~\cite{noh2020efficient, aharonov2023polynomial, PRXQuantum.5.010334, schuster2024polynomial}---a feature that has been exploited in algorithms for computing transport coefficients~\cite{PhysRevB.105.245101}. In addition, present-day experiments can access nonlocal observables that clearly distinguish between isolated and dissipative quantum systems~\cite{PRXQuantum.3.040329, dalzell2024random, ware2023sharp, morvan2023phase}, such as the bit-string distribution after a global measurement of the system~\cite{arute2019quantum, choi2023preparing, bluvstein2024logical}. In Refs.~\cite{ware2023sharp, morvan2023phase} it was pointed out that a first-order phase transition occurs in random circuit sampling if the noise strength $\gamma \to 0$ and the system size $L \to \infty$ limits are taken with $\gamma L$ fixed. If one instead takes the conventional $L \to \infty$ limit at fixed $\gamma > 0$ or $\gamma=0$, the dissipative and unitary systems are in different phases: the former is asymptotically easy to sample from; the latter is hard. 

Recently, Mori~\cite{Mori_2024} (see also Refs.~\cite{Prosen_2002, Prosen_2004, Yoshimura_2024, Znidaric_2024}) considered the spectrum of a generic quantum channel $\mathcal{E}$ describing a chaotic many-body Floquet system weakly coupled (in a generic, local way) to a Markovian bath, and argued that a similarly discontinuous $\gamma \to 0$ limit arises, where $\gamma$ is the coupling to the bath. If $\gamma = 0$ at the outset, unitarity forces the spectrum of $\mathcal{E}$ to lie on the unit circle. However, for any $\gamma > 0$ in the thermodynamic limit, there is a gap in their magnitudes between the steady state eigenvalue and all other eigenvalues; this gap remains nonzero in the $\gamma \to 0$ limit. This singular limit was argued to be analogous to Pollicott-Ruelle resonances in classical chaotic systems~\cite{Pollicott, PhysRevLett.56.405, Prosen_2002, kurchan2009six, dyatlov2015stochastic}. Ref.~\cite{Mori_2024} provided numerical evidence and a heuristic explanation in terms of operator spreading. Numerical evidence also supports the claim that the channel's gap sets the long-time decay rate of local autocorrelation functions; however, given that $\mathcal{E}$ is non-Hermitian for any $\gamma > 0$, the connection between the two rates is not direct. Moreover, observables such as sampling probabilities and out-of-time-order correlators (OTOCs) manifestly cross over from unitary to dissipative behavior on a timescale $\sim \gamma^{-1/2}$, which is far longer than the inverse gap of the channel. Thus, for at least some quantities, non-Hermiticity brings about timescales that are unrelated to the spectrum of $\mathcal{E}$ (cf. Refs.~\cite{trefethen1991pseudospectra, PhysRevResearch.5.033145, rakovszky2023defining}). 

In this work, we analyze the dynamics of weakly dissipative chaotic systems with short-range interactions from the perspective of operator spreading. We do not impose any restrictions on the form of the dissipation except spatial locality. Following Ref.~\cite{Mori_2024}, we consider channels that do not have conservation of energy or any other extensive conserved quantities that will produce gapless transport modes in the $\gamma \to 0$ unitary limit. We present a coarse-grained hydrodynamic description of operator dynamics in the presence of dissipation in terms of operator size distributions~\cite{Schuster_2023}. We find that eigenmodes of the operator-size distribution remain gapped in the weak dissipation limit. For generic space- and time-translation invariant channels, we use this hydrodynamic perspective to argue that the spectrum is gapped, and to characterize its low-lying eigenmodes. While this analysis relies on a standard phenomenology of chaotic operator growth, we are able to microscopically derive the relevant hydrodynamic equations for random unitary circuits subject to weak dissipation. In this case, there is no eigenvalue spectrum, but there is a \emph{Lyapunov spectrum}, whose properties we discuss. We discuss how these spectra relate to the dynamics of correlation functions and OTOCs. Finally, we note that a spectral phase transition can occur in one dimension in the $\gamma \to 0$ limit: the gap remains nonzero across this transition, but changes nonanalytically as the parameters of the unitary dynamics are tuned. This spectral transition is an operator endpoint trajectory-unbinding transition in the dominant contributions to long-time autocorrelation functions~\cite{Nahum_2022}. 

\emph{Models}.---We consider quantum channels $\mathcal{E}$ that are spatially (quasi-)local, with a tunable dissipation rate $\gamma$. Two examples of such channels are: (i)~a time-dependent Lindbladian, of the form $\mathcal{L}(\rho) = -i[H(t), \rho] + \gamma \sum_i (O_i \rho O_i^\dagger - \frac{1}{2} \{ O_i^\dagger O_i, \rho \})$, where $H(t)$ is a generic local Hamiltonian, and $O_i$ is a local (for concreteness, single-site) operator; and (ii)~a brickwork circuit with dissipation, which one can write as $\mathcal{E}_{t}(\rho) = U_{t}\lb \otimes_{i} \mc{D}_{i}\l \rho\r \rb U_{t}$, where $U_t$ is a single repeating unit of a brickwork unitary circuit~\cite{Nahum_OpRUCs, Keyserlingk_OpRUCs} and $\mathcal{D}_i(\rho)$ is a single-site channel that implements arbitrary decoherence at a rate $\gamma$ (and otherwise does nothing). We will always consider the adjoint (``Heisenberg'') action of the channel on observables~\cite{mikeandike}. From this perspective, it does not matter whether the channel is unital, since the adjoint is always a unital map. For simplicity, however, we will take each $\mathcal{D}_i$ to have a unique steady state (unlike, e.g., the dephasing channel). In the case where the Hamiltonian is time-periodic (or the channel repeats identically) one can talk about eigenvalue spectra. In the more general case where the time-dependence is arbitrary but the ensemble of Hamiltonians or unitaries is time-invariant we can instead talk about Lyapunov spectra. Although these cases are formally different, the physics we discuss is largely insensitive to their differences.

\emph{Operator spreading}.---A Pauli string is a tensor product of Pauli operators ($I, X, Y, Z$) on each site of the system: e.g., on a three-site system, $X_1 \otimes I_2 \otimes X_3 \equiv XIX$ is a Pauli string. The support of a Pauli string is the set of sites on which it is not the identity. For a one-dimensional system with open boundary conditions, we can define a left (right) endpoint $x_{L(R)}(S)$ for each Pauli string $P_S$, as the leftmost (rightmost) site on which the string is not the identity, $I$. The $(4^L-1)$ non-identity Pauli strings in a system of size $L$ form an orthonormal basis (in the Frobenius inner product) for operators, in terms of which we expand our traceless operator as $O = \sum_S a_S P_S$, $|a_S| > 0$. We define the right-weight of $O$ as $n(x) = \sum_{\{ S | x_R(P_S) = x\}} |a_S|^2$. (The left-weight can be defined similarly.) As $O$ evolves, the right-weight also evolves, to $n(x,t)$. For concreteness, we consider chains of qubits; the generalization to qudits is straightforward.

Now we develop an evolution equation for $n(x,t)$. To simplify, we restrict ourselves to systems with open boundary conditions, placing our initial $O$ on the single leftmost site $x=1$. Under unitary evolution $O$ spreads to the right, while $x_L$ remains near $x_L=1$. After evolving for time $t$, a typical Pauli string in $O$ is supported on $\sim v_B t$ sites, where $v_B$ is the butterfly speed~\cite{Nahum_OpRUCs, Keyserlingk_OpRUCs}. $O$ can only grow at the edges of its support because of the unital property. In general, the right endpoint $x$ undergoes a random walk with a rightward bias. For random unitary circuits, the parameters of this random walk can be explicitly computed~\cite{Nahum_OpRUCs, dahlsten2007emergence, PhysRevA.78.032324}, and it is a purely Markovian process. More generally, there could be memory effects: however, when $x \gg 1$ and the system is quantum chaotic, the coarse-grained dynamics of the endpoints is expected to be Markovian on physical grounds, since the complex evolution of the ``interior'' of the operator effectively dephases its endpoints~\cite{PhysRevB.107.094311}. 

The effects of local dissipation can be incorporated as follows~\cite{Schuster_2023}: a Pauli string supported on $\ell$ sites decays at a rate $\sim \ell\gamma$. Once again, we can treat the interior of each Pauli string as ergodic, so that (up to a rescaling) the weight $n(x)$ decays at a rate $\gamma x n(x)$. Putting these considerations together, we arrive at the equation
\beq\label{markov}
\dot{n}(x) = w_+ n(x-1) + w_- n(x+1) - (w_- + w_+ + \gamma x) n(x),
\eeq
where $w_+>w_-$ and we have suppressed the time indices~\footnote{Random brickwork circuits follow a discrete-time version of Eq.~\eqref{markov}, namely $n_{x, t+1} = w_+ n_{x-1,t} + w_- n_{x+1,t} + (1-w_+-w_- - \gamma x) n_{x,t}$.}. Eq.~\eqref{markov} describes random circuits in which a gate is randomly applied to each bond at some rate. It can more generally be viewed as an approximation that captures the coarse-grained operator dynamics in generic chaotic systems. In general, the dynamics of small operators will have system-specific features which we can capture approximately by letting $w_\pm, \gamma$ differ from their bulk values in some region $x \leq x_0$, where $x_0$ does not scale with $\gamma$. Finally, for a system of size $L$ there are hard-wall boundary conditions so $1\leq x\leq L$.

\begin{figure}[tb]
    \begin{center}
    \includegraphics[width=0.47\textwidth]{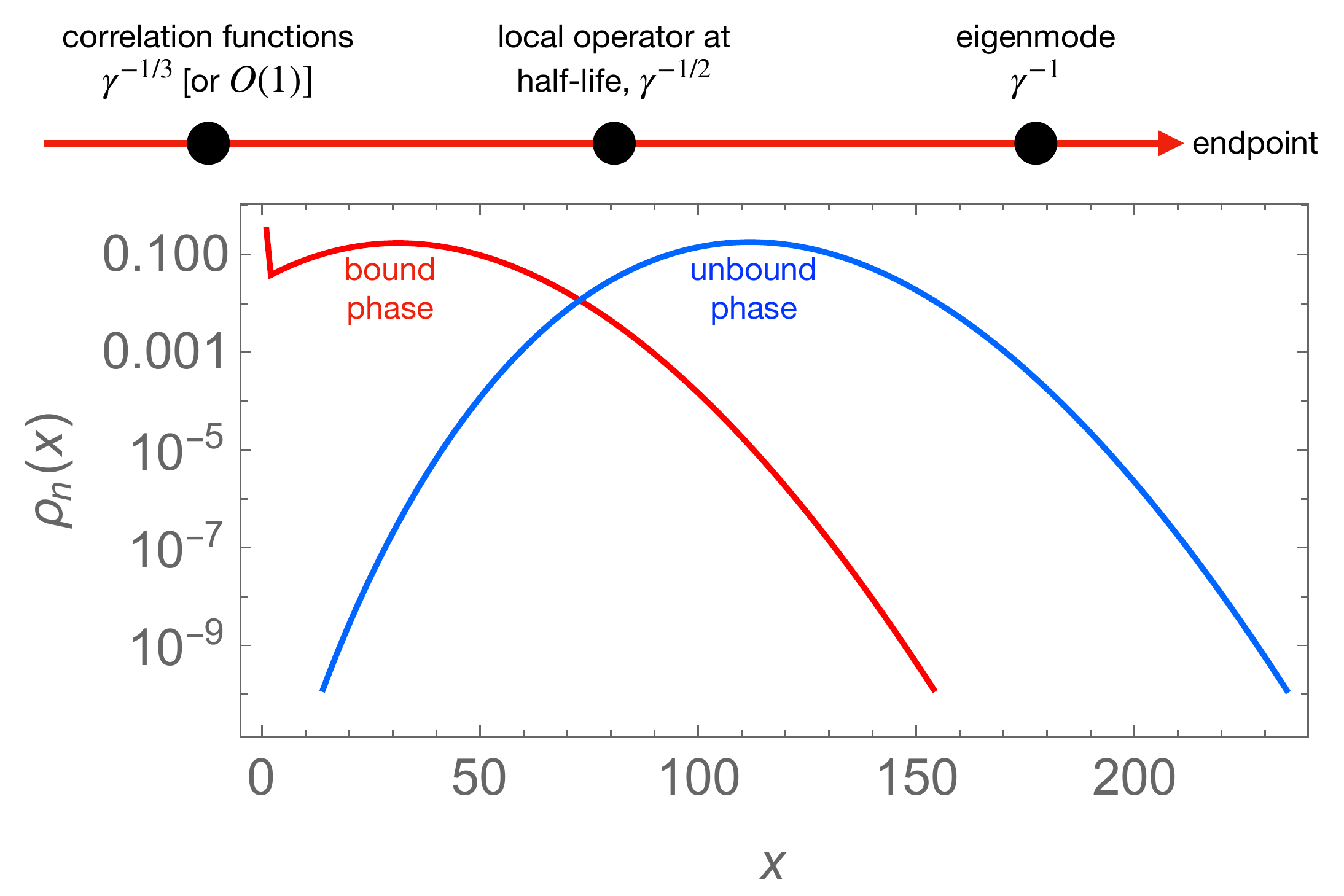}
    \caption{Upper panel, from right to left: Characteristic operator lengths in the distributions produced by the slowest eigenmode, an initially local operator when it has lost half its norm to the bath, and in the dominant paths contributing to long-time correlation functions. As $\gamma \to 0$ these become parametrically distinct.  Lower panel: example profiles of the lowest-lying mode of Eq.~\eqref{markov} for $\gamma = 0.01 w_-, w_+ = 4 w_-$ (corresponding to random unitary circuits on qubits), in the bound phase (red) and unbound phase (blue).  To make the bound phase, the ``hopping'' rates $w_{\pm}$ between lengths $x=1$ and $2$ are reduced by a multiplicative factor of $g=1/10$ relative to larger $x$.  The binding transition at $\gamma=0$ occurs at $g_c=1/3$ for qubits.}
    \label{fig1}
    \end{center}
\end{figure}

\emph{Spectrum and eigenstates}.---Eq.~\eqref{markov} can be recast as an eigenvalue problem for a non-Hermitian operator $M \equiv \sum_{x,x'} M_{x,x'} \ket{x} \bra{x'}$, where $x$ is a positive integer. $M$ can be made Hermitian by the similarity transform $\tilde{M} = T^{-1} M T$, where $T$ is a diagonal matrix.  In the bulk, $T = e^{a x / 2} \delta_{x,x'}$ with $e^{a} = w_+/w_-$. For $x\leq x_0$ where the $w_\pm$ vary, $T$ is adjusted accordingly. Additionally, $\tilde M = \tilde M_0 - \gamma x \delta_{x,x'} - \Lambda \delta_{x,x'}$ in the bulk, where $\tilde M_0$ is the generator for a symmetric random walk with hopping rate $w = \sqrt{w_+ w_-}$, and $\Lambda = (\sqrt{w_+} - \sqrt{w_-})^{2}$. $\Lambda$ gives a constant shift to the eigenvalues of $\tilde M$  for bulk eigenstates, and is the origin of the nonzero Mori gap in the bulk. 

We now turn to the spectrum of $\tilde M$, ignoring (for the moment) any potential subtleties from the short-operator behavior at small $x$. For small $\gamma$ we take a continuum limit on the Hermitian operator $\tilde{M}$. This yields the following continuum differential equation for the bulk eigenfunctions $\psi$ and eigenvalues $-\lambda$ in the transformed coordinates: 
\beq\label{continuum}
(-\lambda + \Lambda) \psi(x) = w \psi''(x) - \gamma x \psi(x). 
\eeq
This equation is supplemented with a Dirichlet boundary condition $\psi\l 0 \r = 0$, quantizing the eigenvalues $-\lambda_{n}$. The eigenstates are Airy functions, 
\beq\label{airy}
\psi_{n}(x) = \mathrm{Ai}\lb\l \frac{\gamma}{w} \r^{1/3} \l x - \frac{\lambda_{n} - \Lambda}{\gamma}\r\rb.
\eeq
Accordingly the slowest (Mori) mode has eigenvalue $-\lambda_{1}$ with  $\lambda_{1} - \Lambda \sim \gamma^{2/3} w^{1/3}$. The ``excited'' eigenvalues of this coarse-grained, continuous-time approximation are spaced as $\lambda_{n} - \Lambda \sim (w \gamma^2 n^2)^{1/3}$, although for small $\gamma$ each such mode represents many eigenmodes of the channel $\mathcal{E}$ so the gap structure beyond $\lambda_1$ is likely peculiar to this approximation. 
While the similarity transformation allowed us to take a ``safe'' continuum limit, the nature of these eigenmodes is more transparent in the original coordinates. In these coordinates, the continuum eigenmodes take the form $\phi_n(x) \equiv e^{ax/2} \psi_{n}(x)$, where $\psi_{n}(x)$ is given in~\eqref{airy}. Several remarks are in order. (1)~The eigenmodes $\phi_n(x)$ are not orthogonal to one another. Their real-space profiles are determined by balancing the exponential growth term against the large-argument decay of the Airy function, $\mathrm{Ai}(z) \sim \exp(-\frac{2}{3} z^{3/2})$. Thus each $\phi_n(x)$ is peaked at $x\sim w/\gamma$. 
(2)~If we take $\gamma = 0$ at the outset, in a system of finite length $L$, the steady state eigenmode is then exponentially localized at $x = L$. This is just the observation that for $\gamma=0$ generic operators spread over the entire system.  For $\gamma>0$, this mode localized near $x=L$ decays at rate $\sim \gamma L$, so by instead taking the $L\to\infty$ limit first, we move this mode to arbitrarily fast decay.
(3)~We considered a one-sided growth process, corresponding to an operator that was initially at one end of the system. The generalization to operators that start in the bulk of the system is fairly straightforward and gives bulk modes with twice the decay rate, because they have two endpoints that are in the bulk of the system.

\emph{Bound states}.---So far, we treated the ``bulk'' of Eq.~\eqref{markov} as terminating at $x = 0$ with simple Dirichlet boundary conditions. In general, however, the dynamics of small-size operators will be different from that in the bulk. An example is when the unitary dynamics has a nearly conserved quantity which decays at a rate $\Gamma$. In this case, Eq.~\eqref{markov} can still be transformed into Hermitian form, but in that ``frame'' it will have an attractive potential at $x \leq x^*$, where $x^*$ is a scale that does not depend on $\gamma$ or $L$. If this attractive potential is strong enough, it can create a bound state of $\tilde{M}$; there is a threshold for forming a bound state because of the Dirichlet boundary condition at $x = 0$. The distinction between bound and scattering states is not sharp for any $\gamma > 0$ because of the confining potential, but it becomes sharp as $\gamma \to 0$. This bound state formation problem was considered in Ref.~\cite{Nahum_2022}; we now discuss its implications for the spectrum of $\mathcal{E}$. 

First, we consider the behavior when there is a ``deep'' bound state with decay rate $\Gamma\ll\Lambda$.  In the Hermitian frame, the bound state decays for small $x$ as $e^{-x/\xi}$ for some $\xi$. At larger distances, this exponential decay becomes super-exponential when $\gamma>0$ because of the linear confining potential. Translating this behavior back into the original coordinates, one finds that the distribution of Pauli string lengths in this bound state is peaked at $x \sim \Gamma/\gamma$. Next, we consider the behavior near the binding threshold.  We tune through this transition by varying some bare parameter $g$ of the small $x$ dynamics.  This could be the rate at which a length-1 operator becomes length-2.  As one tunes $g$ in the bound phase near the threshold with $\gamma=0$, the binding ``energy'' evolves as $(\Lambda-\Gamma) \sim (g_c - g)^2$, and its size scales as $\xi \sim 1/(g_c - g)$.  In the unbound phase, the gap $\Gamma$ is set by the bulk gap $\Lambda$, so $\Gamma=\Lambda$.  At nonzero $\gamma$ the transition between the bound and unbound phases is rounded out on a length-scale set by the eigenstates in the Airy potential (when $|g_c - g| \sim \gamma^{1/3}$).

In higher dimensions $d$ the lowest-lying modes are always bound states~\cite{Nahum_2022}. Therefore for $d>1$, the gap $\Gamma$ is always set by the rate at which order-one-sized operators grow into larger operators.

\emph{Operator dynamics.}---The correlation functions of local observables are generated by Eq.~\eqref{markov} acting on the state $\ket{1}$, which denotes the distribution with all its weight on length-one operators on the leftmost site.  After evolving for a time $t$, for $\gamma=0$ the operator endpoint distribution is centered at $x=v_B t$ and has a width $\sqrt{D t}$.  To first order in $\gamma$, the number of dissipation events in the history of the typical operator endpoint is $\sim \gamma v_B t^2$, so the half-life of the operator scales as $1/\sqrt{\gamma v_B}$: the operator evolves to a size $\sim \sqrt{v_B/\gamma}$ before it decays into the bath. These length and time scales are what sets the dynamics of quantities like the OTOC~\cite{Schuster_2023}. They are not directly related to the slow eigenmodes of the channel: by the time an operator spreads out over a length-scale $O(1/\gamma)$, the characteristic scale of the slow eigenmodes, its norm will be exponentially small in $1/\gamma$. Thus, on the timescales when correlations or OTOCs decay, an initially local operator is not well-described by its projection onto the lowest-lying eigenmode. This observation is not surprising given the non-Hermiticity of Eq.~\eqref{markov}. However, it does raise an apparent puzzle. Mori~\cite{Mori_2024} numerically found that the decay rate of autocorrelation functions on $O(1)$ timescales matches the gap of $\mathcal{E}$. Why should this be, given that the Heisenberg-evolved operator at these timescales looks nothing like the slowest eigenvectors of $\mathcal{E}$?

We first explain why these rates match from the solution of Eq.~\eqref{markov}, and then provide a more general heuristic picture. 
To explicitly map the dynamics of correlation functions to that of operator sizes, we consider ensembles of Haar-random circuits $\{ U \}$ \cite{Collins_2002, Collins_2006, Nahum_OpRUCs, Keyserlingk_OpRUCs} on qudits (of dimension $q$). 
The mean-square value (averaged over circuits) of the autocorrelation function of a generic length-one traceless operator $O$ on the leftmost site is $C^2(t) \equiv \mathbb{E}_U(({\rm Trace}\{O(t) O(0)\})^2)$. This is proportional to the return probability of the operator endpoint: $C^2(t)=\langle n(1,t)  \rangle_c/(q^2-1)$, where $\langle \cdot \rangle_c$ denotes averaging over realizations of the Markov process~\eqref{markov}. In the position basis, therefore, $C^2(t) = \langle 1 | T^{-1} e^{\tilde M t} T | 1 \rangle/(q^2-1)$.  We can normalize $T$ so that $T |1 \rangle = |1\rangle$, so up to a constant $C^2(t) \sim \langle 1 | e^{\tilde M t} | 1\rangle$.  When the slowest eigenmode of $\tilde{M}$ is a bound state with eigenvalue $-\Gamma$, this state has an order-one overlap with $|1\rangle$ and we have $C^2(t) \sim e^{-\Gamma t}$.  When it is unbound, so has gap $\Gamma=\Lambda$, then we have $C^2(t)=e^{-\Lambda t}G(t)$, where $G(t)$ is the Green's function of $(\tilde{M}+\Lambda)$ for leaving from and returning to the leftmost site. To leading order, therefore, mean-square correlation functions of edge operators decay exponentially with the rate $\Gamma$ at long times~\footnote{This long-time exponential decay of autocorrelations can be delayed at short time for operators of length $\ell$, with the delay being by time up to $\sim (\ell/v_B)$; see H. Kim, {\it et al.}, Phys. Rev. E {\bf 92}, 012128 (2015).}, with subleading corrections that differ among the bound and unbound phases and the critical point between them.

In terms of the solution of Eq.~\eqref{markov}, correlation functions are special because both the initial and final boundary conditions are $\ket{1}$, which transforms trivially under $T$. By contrast, quantities like the OTOC are dominated by large operators, whose profiles are totally different in the original and Hermitian frames. 
We can also offer a more intuitive explanation of why the decay rates of correlation functions set the gap. The density of small operators decays primarily because they grow larger; very large operators, instead, primarily decay into the bath. To make an eigenmode of the operator size distribution, the two decay rates must balance, and they are equal to the gap $\Gamma$. This balance is ensured because the eigenmode has a size $\ell$ such that $\Gamma = \ell \gamma$.  Small operators evolve and decay to become operators that have weight $\sim e^{-\Gamma t}$ of being small operators at long times (causing local correlation functions to decay that way). 

Chaotic dual unitary circuits~\cite{PhysRevX.9.021033, PhysRevB.100.064309, PhysRevLett.123.210601} offer an instructive limit. In these circuits, all initially local operators eventually grow deterministically at every time step. Therefore, local autocorrelation functions identically vanish after some finite time, and the associated long-time decay rate is infinite. The gap $\Lambda$ is also extensive: any distribution of operator endpoints that is supported at $x\ll L$ will eventually move outward at every time step and thus cannot be an eigenmode for small $\gamma>0$, at least until it reaches the end of the system. Therefore, the dynamics of operator growth does not admit any eigenstates with a decay rate that remains finite as $L \to \infty$, even for small $\gamma>0$.

The discussion for space-time random circuits above was for the mean-square correlation function. In general this will not match the correlation function in a typical circuit. For example, in the phase with a bound state, the probability for an operator to survive at size $1$ for a time $t$ is a product of independent random probabilities, and has a broad log-normal distribution. Spacetime randomness also affects the critical properties of the unbinding transition~\cite{Nahum_2022}. The mean-square correlation function is dominated by realizations in which correlations decay anomalously slowly; thus, its decay rate $\Gamma$ is lower than that of a typical circuit. By Markov's inequality, the probability of a circuit with an apparent decay rate $\Gamma'$ after time $t$ is bounded as $P\left[\langle O(t) O(0) \rangle^2 \geq e^{-\Gamma' t}\right] \leq e^{-(\Gamma - \Gamma') t}$, so decay rates with $\Gamma' < \Gamma$ are vanishingly rare.

\emph{Generic chaotic channels}.---We now consider a generic chaotic channel $\mathcal{E}$ that is translation invariant in discrete space and time. When $\gamma = 0$, the spectrum of $\mathcal{E}$ all lies on the unit circle. For $\gamma > 0$, we expect the entire spectrum (except for the unique steady state) to live inside a disc of radius $ e^{-\Gamma}$, where the time unit is taken to be one application of the channel $\mathcal{E}$, and the gap $\Gamma$ is well approximated for small $\gamma>0$ by the gap of~\eqref{markov}. 
As discussed above, Eq.~\eqref{markov} is believed to be a good approximation for the dynamics of large operators in general chaotic systems, so for small $\gamma>0$ the slowest eigenmode of the operator length distribution in Eq.~\eqref{markov} will approximate the slowest eigenmode of $\mathcal{E}$. But note that $\mathcal{E}$ has $4^L$ eigenvalues while the Markov process on the endpoint distributions has only $L^2$ eigenvalues, so the correspondence between these eigenmodes must be many-to-few, and is an interesting open question to further explore. 

\emph{Discussion}.---We have presented a description of the low-lying eigenmodes of generic chaotic quantum channels with only short-range interactions in terms of a coarse-grained picture of operator dynamics in open quantum systems. Since it works with eigenvectors in the adjoint (``Heisenberg'') picture, our analysis does not distinguish between unital and non-unital channels. For random unitary circuits, our coarse-grained theory can be microscopically derived, and lower bounds the decay rate of autocorrelations in typical circuits. For generic chaotic systems, our treatment is phenomenological, but captures the coarse-grained structure of eigenvectors. Our results explain why the decay of local autocorrelation functions matches the decay of the very large operators that constitute the dominant eigenmodes. Including conservation laws (and the resulting hydrodynamic modes) is a natural extension of our results. It would also be interesting to adapt our analysis to classical chaotic systems with spatial locality (e.g., classical spin chains) and show a direct relation between operator sizes and the canonical Pollicott-Ruelle resonances that occur in these systems.

The singular $\gamma \to 0$ limits in sampling hardness for noisy random circuits~\cite{ware2023sharp, morvan2023phase} and in the gap of $\mathcal{E}$ are closely related. In Ref.~\cite{ware2023sharp} the problem of sampling from a noisy channel was mapped onto an Ising model in a longitudinal magnetic field of strength $\gamma$. Setting $\gamma = 0$ gives a doubly degenerate ground state, but for any $\gamma > 0$ in the thermodynamic limit, the ground state is unique with an $O(1)$ gap independent of $\gamma$. The gap of $\mathcal{E}$ corresponds to the $O(1)$ cost to create domain walls, while the longitudinal field penalizes a minority domain of size $\ell$ by an ``energy'' cost $\sim \gamma \ell$, so domain walls experience a linear confinement that corresponds to the linear potential in Eq.~\eqref{continuum}. Relating computational hardness transitions to spectral properties of the channel remains a direction for future work.

\emph{Related Work}.--- We recently became aware of a parallel and independent work by Carolyn Zhang, Laimei Nie, and Curt von Keyserlingk, Ref.~\footnote{Carolyn Zhang, Laimei Nie, and Curt von Keyserlingk, Thermalization rates and quantum Ruelle-Pollicott resonances: insights from operator hydrodynamics.}, which will appear in the same arXiv posting as this work. Our conclusions seemingly agree where they overlap.

\begin{acknowledgments} \emph{Acknowledgments}.---
    The authors thank Ignacio Cirac, Curt von Keyserlingk, Jorge Kurchan, Tibor Rakovszsky, Rahul Trivedi, Sagar Vijay, and Peter Zoller for helpful discussions. 
     J.A.J. was supported by the National Science Foundation Graduate Research Fellowship Program under Grant No. DGE-2039656 for the duration of this work.  D.A.H. was partially supported by U.S. National Science Foundation QLCI grant OMA-212075. S.G. was partially supported by NSF 
     QuSEC-TAQS OSI 2326767. 
     Any opinions, findings, and conclusions or recommendations expressed in this material are those of the authors and do not necessarily reflect the views of the National Science Foundation.
\end{acknowledgments}
\bibliography{main_post.bbl}

\end{document}


\newcommand{\grad}{\vec{\nabla}}
\newcommand{\curl}{\vec{\nabla}\times}
\newcommand{\divergence}{\vec{\nabla}\cdot}
\renewcommand{\l}{\left(}
\renewcommand{\r}{\right)}
\newcommand{\lb}{\left[}
\newcommand{\rb}{\right]}
\newcommand{\lcb}{\left\{ }
\newcommand{\rcb}{\right\} }
\newcommand{\bra}{\left<}
\newcommand{\ket}{\right>}
\newcommand{\lv}{\left|}
\newcommand{\rv}{\right|}
\newcommand{\mbf}[1]{\mathbf{\mathrel{#1}}}
\newcommand{\mc}[1]{\mathcal{\mathrel{#1}}}
\newcommand{\up}{\uparrow}
\newcommand{\down}{\downarrow}
\newcommand{\pI}{\mathbb{I}_{2}}
\newcommand{\pX}{\mathbf{X}}
\newcommand{\pY}{\mathbf{Y}}
\newcommand{\pZ}{\mathbf{Z}}
\newcommand{\npI}{\hat{\mathbb{I}}_{2}}
\newcommand{\npX}{\hat{\mathbf{X}}}
\newcommand{\npY}{\hat{\mathbf{Y}}}
\newcommand{\npZ}{\hat{\mathbf{Z}}}
\newcommand{\dbar}{\mathchar'26\mkern-12mu d}
\renewcommand{\log}[1]{\, {\rm Log} \left[ \mathrel{#1} \right] }
\newcommand{\titlemath}[1]{\texorpdfstring{$\mathrel{#1}$}{TEXT}}
\renewcommand{\=}[1]{$\mathrel{#1}$}
\newcommand{\tr}[1]{{\rm Tr}\lb\mathrel{#1}\rb}
\newcommand{\ptr}[2]{{\rm Tr}_{#2}\lb\mathrel{#1}\rb}
\newcommand{\svn}[1]{\mc{S}_{\rm vN}^{\rm \mathrel{#1}}}
\newcommand{\mbs}[1]{\boldsymbol{#1}}
\renewcommand{\mod}[1]{\, {\rm mod} \; #1}
\newcommand{\oft}{\l  t \r}
\newcommand{\vb}{v_{\rm B }}

\preprint{APS/123-QED}

\title{Supplementary Material: Spectral gaps of local quantum channels in the weak-dissipation limit}

\author{J.~Alexander~Jacoby}
\affiliation{Department of Physics, Princeton University, Princeton, New Jersey 08544, USA}

\author{David~A.~Huse}
\affiliation{Department of Physics, Princeton University, Princeton, New Jersey 08544, USA}

\author{Sarang~Gopalakrishnan}
\affiliation{Department of Electrical and Computer Engineering, Princeton University, Princeton NJ 08544, USA}

\date{\today}

\maketitle
\onecolumngrid

\section{Bulk Microscopics}
\label{sect:Microscopics}

\subsection{Endpoint Markov Process for RUCs: Bulk Eigenmodes}
\label{sect:endpoint_markov}
Per the main text, in a Random Unitary Circuit (RUC) operator spreading can be understood in terms of biased random walks of operator string endpoints \cite{Nahum_OpRUCs, Keyserlingk_OpRUCs, Collins_2002,Collins_2006}. We consider a circuit with on-site Hilbert space dimension $q$ (qudits) and begin with an operator string that has its left/right endpoint on $x_{L/R}$ (taken to be far from the boundaries of the system), where a two-qudit Haar-random gate of the brickwork circuit is applied. A two-site unitary gate randomizes among the $q^{4}-1$ linearly independent and traceless two-site operators supported on the two sites to which the gate is applied (one of which is the string endpoint site). Of these operators $q^{2}-1$ are the identity on the ``forward'' site (the site further from the rest of the string's support). Randomization by the unitary results in a probability $p = \l q^{2} -1   \r/\l q^{4}-1 \r = 1/\l q^{2}+1\r$ that the forward site is an identity after the gate is applied (and probability $q= 1-p = q^2 / \l q^2 + 1 \r $ that it is non-identity).

\begin{figure}[h]
    \centering
    \includegraphics[width = 0.5 \linewidth]{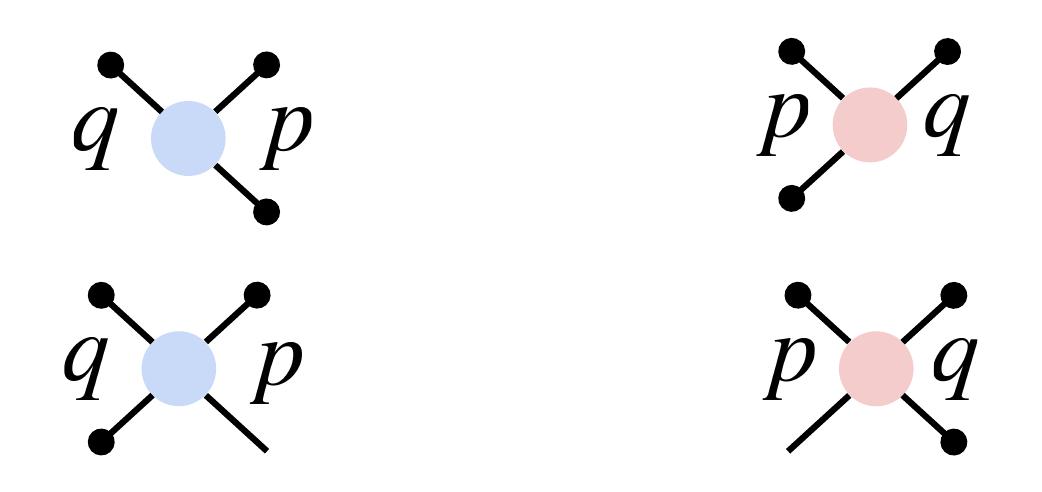}
    \caption{Isolated left (blue) and right (red) string endpoints under continuous time RUC dynamics. Colored circles are gates. Lines with bulbs at their termini are incoming and outgoing operator string endpoints. Outgoing string endpoints are labeled by the probability of their outgoing position with respect to the gate given their incoming position with respect to the gate. The upper configurations show gates which allow an operator string to expand, whereas the bottom configurations show gates which allow an operator string to contract. Each configuration above occurs with equal rate $r$.}
    \label{fig:cont_time_endpoint_jumps}
\end{figure}

We would like to treat the continuous time RUC for operators that have both endpoints in the bulk, and therefore we consider the application of gates on each dual-lattice site as a Poisson process with rate $r$. We define $x_{L/R}$ to take values on the lattice, rather than the dual-lattice (a common convention in brickwork circuits). There are four ways in which gates can act on the endpoint, which are summarized in Fig.~\ref{fig:cont_time_endpoint_jumps}. 

If gates arrive with rate $r$ on each dual-lattice site and each gate moves the string endpoint by one lattice unit, we find that the left and right endpoints independently obey the equations
\begin{eqnarray}
    \dot{n}_{L}\l x_{L}, t \r &=& r\l \frac{q^{2}}{q^{2}+1}n_{L}\l x_{L} + 1, t \r +  \frac{1}{q^{2}+1}n_{L}\l x_{L} - 1, t \r - n_{L}\l x_{L} , t \r\r \nonumber \\
    \dot{n}_{R}\l x_{R}, t \r & = & r \l \frac{1}{q^{2}+1}n_{R}\l x_{R}+1, t \r + \frac{q^{2}}{q^{2}+1}n_{R}\l x_{R} -1, t \r - n_{R}\l x_{R}, t \r\r.
\end{eqnarray}
The separation of the left and right endpoint equations requires two assumptions: (1) the operator is inserted in the bulk of the system far away from any boundaries (2) we can safely neglect endpoint interactions (the endpoints are not bound). We would like to separate out the size distribution of an operator, as the dissipation is diagonal in the size basis (see Section~\ref{sect:size_and_diss}). We therefore define the center of mass coordinate $Y = \l x_{L}+x_{R}\r /2$ and relative coordinate $X=  x_{R} - x_{L} >0$, which is just the size of the string. The attractive interaction can be accounted for directly in the relative coordinate equation (e.g.,  \cite{Nahum_2022} for brickwork circuits). The Markov process equations in these coordinates separate as,
\begin{eqnarray}
    \dot{n}_{\rm REL}\l X ,t  \r &=& 2r\l \frac{q^{2}}{q^{2}+1}n_{\rm REL}\l X-1, t \r + \frac{1}{q^{2}+1} n_{\rm REL}\l X+1, t \r-n_{\rm REL}\l X, t \r\r \nonumber \\
    \dot{n}_{\rm COM}\l Y, t \r &=& r\l n_{\rm COM}\l Y +1/2, t \r + n_{\rm COM}\l Y -1/2, t \r - 2 n_{\rm COM}\l Y , t \r\r
\end{eqnarray}
The continuum limit on the COM distribution must be taken somewhat carefully in $2Y$ rather than $Y$, and yields the unbiased diffusion equation for $n_{\rm COM} $ with constant $D_{\rm COM} = r/4$. The continuum limit for $n_{\rm REL}$ is the same as described in the main text and detailed in Section~\ref{sect:cont_lim}. In terms of the parameters of the main text and Section~\ref{sect:eigenmode_details}, it has $a = 2\log{q}$, $w = 2rq/\l q^{2} +1 \r$, and $\Lambda = 2r \l q - 1\r^{2}/\l q^{2} + 1\r$. The eigenmodes are given in terms of these parameters by Eq.~\ref{eqn:eigenmodes}. This $\Lambda$ is the $\gamma \to 0 $ gap for operators with both endpoints in the bulk in the case where there is no bound state. The case considered in the main text is where one endpoint is at the edge of the system, and then the bulk gap is instead $\Lambda = r \l q - 1\r^{2}/\l q^{2} + 1\r$ (also in the $\gamma \to 0$ limit).  If there is a bound state with gap $\Gamma$, this can be a bound state at the edge of the system or a bound state in the bulk of the system, which in general will have different gaps.

\subsection{Diffusion Frame Transformation and Continuum Limit}
\label{sect:cont_lim}
In the continuous time setting, we to aim Hermitize the matrix
\begin{equation}
     M = 
    \begin{pmatrix}
        \ddots & &\vdots & & \reflectbox{$\ddots$}  \\
        &- w_{+} - w_{-} & w_{-} &  &  \\
        \dots& w_{+} & - w_{+} - w_{-}  & w_{-} &  \dots \\
        &  &  w_{+} &- w_{+} - w_{-} &   \\
        \reflectbox{$\ddots$}& & \vdots & & \ddots
    \end{pmatrix} - \gamma X.
    \label{eqn:general_markov}
\end{equation}
Since $T = e^{aX/2} $, where $e^{a} = w_{+}/w_{-}$, is diagonal it commmutes with the diagonal elements of $M$ so 
\begin{eqnarray}
    \tilde{M} = T^{-1}MT &=& 
    \begin{pmatrix}
        \ddots & &\vdots & & \reflectbox{$\ddots$}  \\
        &0 & \l w_{-} w_{+}\r^{1/2} &  &  \\
        \dots& \l w_{-} w_{+}\r^{1/2} & 0  & \l w_{-} w_{+}\r^{1/2} &  \dots \\
        &  &  \l w_{-} w_{+}\r^{1/2} & 0 &   \\
        \reflectbox{$\ddots$}& & \vdots & & \ddots
    \end{pmatrix}
    -\l w_{-}+ w_{+}\r I - \gamma X .
    \label{eqn:Hermitized_Markov}
\end{eqnarray}
Then, $2\l w_{-} w_{+}\r^{1/2}$ can be subtracted along the diagonal of the matrix and added to the constant term to get 
\begin{equation}
    \tilde{M} \xrightarrow[]{\rm continuum} \  w\nabla^{2} -\l w_{-} + w_{+} -  2\l w_{+}w_{-}\r^{1/2}\r - \gamma X = w \nabla^{2}  - \gamma\l X + \Lambda/\gamma \r
    \label{eqn:cont_lim}
\end{equation}
where $w = \l w_{+}w_{-}\r^{1/2}$ and $\Lambda = \l w_{+}^{1/2} - w_{-}^{1/2}\r^{2}$ per the main text.

\subsection{Pauli Weight, Operator Length, and Bulk Dissipation}
\label{sect:size_and_diss}
A circuit with on-site Hilbert space dimension $q$ possesses a generalized Pauli basis of $q^{2}$ operators on each site. Tensor products of these operators, all of which are traceless except for the identity, are known as Pauli strings; $S$ denotes such a string.  In one dimension, the length of a given Pauli string, $X(S)$, is defined as one plus the lattice distance between the furthest two non-identity operators. The unital property of an adjoint channel implies that single-site identity operators remain untouched by the channel.  If a channel with rate $\gamma$ does not selectively dissipate among the traceless, non-identity operators, a random operator on a single site will dissipate with average rate $\gamma_{d} = \l 1 - 1/q^2 \r\gamma$. Assuming the bulk of the operator is random, a typical string $S$ of large length $X$ in the Pauli string expansion of $O\l t\r$ will therefore dissipate with rate $\gamma_{d}X$, at leading order in small $\gamma$ and large $X$.  Formally, the dissipation rate of a string is exactly $\gamma w $, where $w\l S\r$ is defined as the many-body weight of $S$ (the number of non-identity single-site operators in the string). For operator sizes of $O\l1/\gamma\r$, where dissipation will have a substantial effect in the $\gamma \to 0$ limit, the fluctuations in the many body weight at fixed string length are central limiting and subleading in $\gamma$ for RUCs, and presumably also for general chaotic systems with no conserved densities. Though we leave the subscript implicit, the dissipation rate that appears in almost all equations in both the main text and this supplement is the dressed rate, $\gamma_{d}$.

\section{Additional Details of Eigenmodes}
\label{sect:eigenmode_details}

The continuum eigenmodes $\phi_{n}$ are, as in the main text (now written explicitly as a function of the relative coordinate $X = x_{R} - x_{L}$, or operator size),
\begin{equation}
    \phi_{n}\l X  \r = e^{aX/2}\mathrm{Ai}\lb \l \frac{\gamma}{w}\r^{1/3}\l X - \frac{\lambda_{n} - \Lambda}{\gamma} \r\rb = e^{aX/2} \psi_{n}\l X\r
    \label{eqn:eigenmodes}
\end{equation}
with $\lambda_{n} = \Lambda - a_{n} \l w\gamma^{2}\r^{1/3}$, where $\lcb a_{n}\rcb $ are the strictly negative zeros of the Airy function \cite{DLMF}. These eigenmodes can correspond either to the bulk or edge eigenmodes according to the choice of parameters. For continuous time RUCs, for instance, the relationships between the edge and bulk mode parameters are $w_{\rm bulk} = 2w_{\rm edge}$, $\Lambda_{\rm bulk} = 2\Lambda_{\rm edge}$, and $a_{\rm bulk } = a_{\rm edge}$. As in the main text, we note that the eigenvalues are of the Airy approximation itself and their relationship to the eigenvalues of the channel, $\mc{E}$, is likely nontrivial. $\phi_{n}$ can be maximized to a parameterically good approximation in the $\gamma \to 0$ limit by noting that $\phi'_{n}\l X \r  = \l a/2 +\partial_{X}\log{Ai\lb z \rb} \r\phi_{n}\l X\r$ where $z$ is the argument of the Airy function in Eq.~\ref{eqn:eigenmodes}. This yields
\begin{equation}
    X_{\rm max} = \frac{a^{2} w}{4\gamma} - a_{k} \l w /\gamma \r^{1/3}.
    \label{eqn:continuous_maximum}
\end{equation}

Asymptotically, there are some differences between the continuum solution and its discrete counterpart. Measuring $X$ in lattice units to make it integral valued, the discrete solution for $\psi_{n}$ (in the diffusion frame) on site $k\gg w/\gamma$ will go as
\begin{equation}
    -\log{\psi\l X_{k}\r} \sim k\l \log{k} - \log{ew/\gamma}\r.
    \label{eqn:discrete_tails}
\end{equation}
The continuum solution, on the other hand, will go as $- \log{\psi\l X \r} \sim 2\l \gamma/ w \r^{1/2} X^{3/2}/3 - aX/2 $ (up to a constant). Fortunately, the distinction is inconsequential for the Mori eigenvalue, $\lambda_{1}$, which is set in the Hermitian ``large deviations'' frame at scale $\gamma^{-1/3}$.

Formally, one may also directly take a continuum limit on the Markov process equation before shifting to the Hermitian frame. This approach is incorrect as it does not contain the relevant large deviation physics, but it yields some intuitive insight into the eigenmodes. In terms of the standard phenomenological parameters of operator spreading, the butterfly velocity, $\vb$, and diffusion constant, $D$, a pertubative expansion can be developed in $\lambda = \vb/DX_{\rm eq}^{1/2}$ with $X_{\rm eq} = 2 \vb/\gamma$. The zeroth-order solution is a Gaussian of width $\sqrt{X_{\rm eq}}$ centered at $X_{\rm eq}$, with corrections in $\lambda$ that can be written in terms of probabilist's Hermite polynomials. This expansion is discussed in \cite{Rosenbloom_59, Hernandez_10}.

\twocolumngrid

\bibliography{supp_post.bbl}